\documentclass[prl, reprint, nofootinbib, showpacs,preprintnumbers,amsmath,amssymb]{revtex4}
\usepackage{varioref,exscale,latexsym,amsmath,amssymb}
\usepackage{graphicx}
\usepackage{graphicx}% Include figure files
\usepackage{slashed}
\usepackage{dcolumn}% Align table columns on decimal point
\usepackage{bm}% bold math
\usepackage{hyperref}
\usepackage{color}
\usepackage{xcolor}

\newcommand{\beq}{\begin{equation}}
\newcommand{\eeq}{\end{equation}}
\newcommand{\bea}{\begin{eqnarray}}
\newcommand{\eea}{\end{eqnarray}}

\def\lsi{\raise0.3ex\hbox{$<$\kern-0.75em\raise-1.1ex\hbox{$\sim$}}}
\def\gsi{\raise0.3ex\hbox{$>$\kern-0.75em\raise-1.1ex\hbox{$\sim$}}}

\def\beq{\begin{equation}}

\def\eeq{\end{equation}}
\def\beqa{\begin{eqnarray}}
\def\eeqa{\end{eqnarray}}

\begin{document}
\preprint{ACFI-T22-01}

\title{{\bf Non-local partner to the cosmological constant}}

\medskip\

\medskip\

\author{John F. Donoghue${}^{1}$}
\email{donoghue@physics.umass.edu}
\affiliation{
${}^1$Department of Physics,
University of Massachusetts,
Amherst, MA  01003, USA}

\begin{abstract}
I show that quantum corrections due to a massive particle generates a non-local term in the gravitational effective action which is of zeroth order in the derivative expansion, much like the cosmological constant. It carries a fixed coefficient which is very much larger than the cosmological constant, and which cannot be fine-tuned. The interaction is active at scales above the particle's mass. This is of the form $m^4 (\frac1{\Box}R)_x `` \langle x|\log (\Box +m^2) |y \rangle'' (\frac1{\Box}R)_y$, and I discuss the meaning of $ `` \langle x|\log (\Box +m^2) |y \rangle'' $ and other aspects of its interpretation.
\end{abstract}
%\vspace{0.2 in}
%\end{titlepage}
%\setcounter{page}{0}
%\newpage
\maketitle
%\documentstyle[12pt,epsfig]{article}
%\documentstyle[12pt,epsf,epsfig]{article}

%%%%%%%%%%%%%%%%%%%%%%%%%%%%%%%%%%%%%%%%
\section{Local and non-local interactions}
The foundation of General Relativity is general covariance - the invariance of the theory under local changes in the coordinates. One can write an action which respects this symmetry using the curvatures
\beq
S =\int d^4x \sqrt{-g} \left[-\Lambda + \frac1{16\pi G}R +c_1 R^2 +c_2 R_{\mu\nu}R^{\mu\nu}+~...\right]
\eeq
Here $\Lambda$ is the cosmological constant (of mass dimension $m^4$), $R_{\mu\nu}$ and $R=g^{\mu\nu}R_{\mu\nu}$ are the Ricci tensor and scalar curvature, $G$ is Newton's constant (with mass dimension $1/G\sim m^2$) and $c_1,~c_2$ are dimensionless constants. This action has been ordered in the derivative expansion, with the cosmological constant being of zeroth order, the scalar curvature having two derivatives and the curvature-squared terms involving four derivatives. The ellipses indicate terms with yet more derivatives. In applications the derivatives turn into factors of energy and momentum, and therefore the terms with two or more powers of the curvature are negligible at low energy. Using the first two terms in this action, we obtain Einstein's equations with a cosmological constant. Rather famously, the cosmological constant is remarkably small, taking the value $\Lambda \simeq 10^{-47}~{\rm GeV}^4$ if the present accelerated expansion of the universe is due the presence of $\Lambda$.

If we had a complete theory of quantum gravity, we could presumably predict the coefficients in the gravitational action from the fundamental parameters in that theory. In place of that knowledge, all that we really know is that the unknown physics from high energy will described by local terms in the action. This follows from the uncertainty principle which tells us that heavy fields do not propagate far when viewed at low energy. When one treats quantum effects of matter fields, or of the gravitational field itself, one needs to renormalize the parameters in this action. The divergences also are local because they occur at high energy. 

However, we can still make predictions at low energy without knowing the complete quantum theory of gravity\cite{Donoghue:1994dn, Donoghue:2017pgk}. At low energy, gravitons and other light particles can propagate a long distance, which distinguishes them from the local high energy effects. Knowing the low energy degrees of freedom and their couplings are sufficient to calculate these effects. In momentum space, where the calculations are most often performed, this is distinguished by non-analytic behavior, such as $\sqrt{q^2}, ~\log q^2$, which cannot be Taylor expanded in powers of $q^2$ and as such cannot be represented by derivative operators in a local Lagrangian. For example, in the quantum correction to the Newtonian potential due to gravition loops, one calculates the
logarithmic non-analytic term
\beq
{\cal M} = \frac{4\pi GMm}{q^2}\left[1 +\frac{41}{20\pi} Gq^2\log{-q^2}\right]
\eeq
which turns into a long-distance correction when Fourier transformed \cite{Bjerrum-Bohr:2002gqz, Khriplovich:2002bt}
\beq
V(r) =- \frac{GMm}{r}\left[1 +\frac{41}{10\pi} \frac{G}{r^2}\right]
\eeq
The one-loop logarithm yields the leading quantum effect.

The non-analytic and non-local terms can be described by a non-local effective action. The most well-known are in the curvature squared terms, where Barvinsky and Vilkovisky and collaborators \cite{Barvinsky:1990up, Barvinsky:1993en, Avramidi:1990ap, Barvinsky:1985an} have included the logarithmic non-analytic terms using the generic form
\beq
{\cal L}=c_1(\mu) R^2 +c_2(\mu) R_{\mu\nu}R^{\mu\nu} + d_1 R \log (\frac{\Box}{\mu^2}) R +d_2 R_{\mu\nu}\log (\frac{\Box }{\mu^2})R^{\mu\nu}
\eeq
with $d_{1,2}$ being known dimensionless constants calculated at one loop order. Here $\log \Box$ represents generalization of the Fourier transform of $\log q^2$ - it will be described more fully below. Although this is written in a form which appears local, it really represents a non-local action because $\langle x|\log \Box|y\rangle$ is non-local. In particular the correspondence is
\beq
{\cal L} = R \log (\frac{\Box}{\mu^2}) R~ ~~\Rightarrow ~~~S= \int d^4x\sqrt{-g(x)}~d^4y\sqrt{-g(y)}~ ~R(x) ~\langle x|\log (\frac{\Box}{\mu^2})|y\rangle ~R (y)
\eeq
In the derivative expansion, these terms are of fourth order in the derivatives.

The basic result of this paper is that when dealing with loops of massive particles, there are also non-local effects at zeroth order and second order in the derivative expansion. That at zeroth order is
\beqa
{\cal L} =&\frac{m^4}{40\pi^2} \left[\left(\frac1{\Box} R_{\lambda\sigma}\right)\left[\log((\Box+m^2)/m^2)\right]\left(\frac1{\Box}R^{\lambda\sigma}\right) 
 -\frac18\left(\frac1{\Box} R\right)\left[{\log((\Box+m^2)/m^2)}\right]\left(\frac1{\Box}R \right)\right]
\eeqa
from the loop of a massive scalar field. A massive fermion yields $-2$ times this result. The two derivatives in the curvatures are canceled by the $1/\Box$ factors, so that overall this is zeroth order in the derivative expansion. The structure of this is explained below, but in the sense of the derivative expansion this is a non-local partner to the cosmological constant. It comes with a coefficient which cannot be adjusted. 

In most settings the fact that there was a non-local component to the interaction would not be remarkable, as these effects are expected in usual quantum corrections. However, in the case of the cosmological constant there is a special feature. The experimental value of the local cosmological constant is very much smaller than expected, while the non-local partner enters at normal size and cannot be removed. Therefore,  it has an enormous numerical advantage over the tiny cosmological constant. For example considering the top quark in the loop, we have $m_t^4/\Lambda \sim 10^{56}$. However, we will see that this interaction is only applicable at scales above the particle's mass, as the effect of the massive field becomes local below that scale.

\section{Origin of non-local interactions}

One-loop diagrams can all be reduced to functions of momentum times the basic scalar tadpole, bubble, triangle and box diagrams \cite{Passarino:1978jh}. The UV divergent diagrams are the tadpole and bubble. The tadpole has no external momentum running through it, and in dimensional regularization has the form near $d=4$
\beqa
I_1&=&-i\mu^{4-d}\int \frac{d^dk}{(2\pi)^d} \frac1{[k^2-m^2]}   \ \ \nonumber \\
&=& \frac{m^2}{16\pi^2}\left[  \frac1{\epsilon} -\gamma +\log(4\pi) +\log(\mu^2/m^2) +1\right]
\eeqa
with $\epsilon =(4-d)/2$. This is always independent of any external momenta. The massive bubble does depend on the external momentum and has the form
\beqa\label{scalarbubble}
I_2(q) &=&-i\mu^{4-d}\int \frac{d^dk}{(2\pi)^d} \frac1{[k^2-m^2][(k-q)^2-m^2]}   \ \ \nonumber \\
&=& \frac1{16\pi^2}\left[ \frac1{\epsilon} -\gamma +\log(4\pi) +\log(\mu^2/m^2) - J(q^2)\right]
\eeqa
with
\beq\label{Jintegral}
J(q^2) = \int_0^1 dx ~\log\left[\frac{m^2 -x(1-x)q^2}{m^2} \right]   \ \ .
\eeq
When the mass is zero or the momentum become large, one obtains the $\log q^2$ which has been referenced above.

The logarithm in the Barvinsky-Vilkovisky effective action comes from the bubble diagram. While they derive this using the background field method and a non-local version of the the heat kernel  \cite{Barvinsky:1985an},  is easy to reproduce it in a perturbative expansion also \cite{Codello:2012kq, Donoghue:2014yha, Donoghue:2015nba,  Donoghue:2015xla}.  One expands
$g_{\mu\nu}(x) = \eta_{\mu\nu} +h_{\mu\nu}(x)$ and calculates the bubble diagram involving two $h_{\mu\nu}$. One finds
\beq
{\cal M} \sim h_{\mu\nu} h_{\alpha\beta} \left(q^\mu q^\nu q^\alpha q^\beta + ...\right) \left[\frac{1}{\epsilon} +... -\log(-q^2)\right]  \ \ .
\eeq
This can be matched to the possible terms in a covariant effective action.There is a unique connection between the tensor structure of the momentum factors and the leading terms in an expansion in the curvatures, resulting in the identification of the renormalization of the curvature-squared terms and also the $R\log \Box R $ terms.

In flat space we have have the clear identification of the meaning of $\log \Box$. It is a function defined by the Fourier transform of $\log -q^2$, 
\beqa\label{logbox}
L_0(x-y) &=& \langle x|\log \Box/\mu^2 |y\rangle \nonumber \\
&=&\int \frac{d^4q}{(2\pi)^4}e^{-iq\cdot (x-y)}\log ( -q^2-i\epsilon)/\mu^2  \ \ .
\eeqa
For curved spacetime we need a covariant generalization of this. This is not uniquely defined - there are several options which could reduce to the flat space version in the appropriate limit.  This issue is discussed in the appendix. 

In the Barvinsky Vilkovisky formalism, there are also non-local terms following from the one-loop triangle diagram, which involves three factors of the gravitational field \cite{Barvinsky:1993en}. These provide the triple graviton portions of the $R^2$ and $R \log \Box R$ effects previously identified,  as well as some new terms. When these latter are written in covariant form they yield contributions which are
proportional to actions such as $R^2 (1/\Box)R$. Here $1/\Box$ is a propagator
\beq
\langle x|\frac{1}{\Box}|y\rangle= G(x,y)
\eeq
which has covariant versions \cite{DeWitt:1975ys, DeWitt:1967ub, DeWitt:1967uc}. Such interactions are said to be of ``higher order in the curvature''. However they are of the same order in the derivative expansion as are the local $R^2$ terms and the nonlocal $R\log \Box R$ terms, i.e. all are of fourth order in the derivative expansion. 

In practice, the Barvinsky Vilkovisky expansion in the curvature is a weak field expansion rather than a derivative expansion. It however has the great advantage that all factors are written in covariant form, so that it manifests the coordinate invariance of general relativity.

\section{Renormalization of the cosmological constant}

What about the lower order terms in the derivative expansion? The coefficients of the curvature-squared terms are dimensionless. Starting with massless fields one can generate these as their coefficients are just pure numbers and do not require any dimensionful parameters. However, to generate terms with fewer derivatives, one needs a extra mass parameter because the coefficients of such term must carry a dimensionful factor. With loops of a massive field, one generates the renormalization of the scalar curvature term at order $m^2$, and the cosmological constant at order $m^4$ when using dimensional regularization, i.e.
\beqa
\delta \sqrt{-g}\frac{1}{16\pi G} R &\sim& \sqrt{-g}\frac{ m^2 R}{\epsilon} + {\rm non-local} \nonumber \\
\delta \sqrt{-g} \Lambda &\sim& \sqrt{-g}\frac{ m^4}{\epsilon} + {\rm non-local}  \ \ .
\eeqa
Our goal is to calculate the non-local actions in these cases. First we need to understand the renormalization of the local terms. 

Again we can identify the effect by an expansion about Minkowski space. For reference, the weak field expansion of the cosmological term is
\beq
\sqrt{-g}\Lambda = \Lambda(1+\frac12 h^\sigma_\sigma + \frac18 ( h^\sigma_\sigma )^2 -\frac14 h_{\sigma\lambda}h^{\sigma\lambda} +...)
\eeq
We can then identify changes in the cosmological constant through the couplings to the gravitational field $h_{\mu\nu}$, with interactions which do not involve derivatives.

For a minimally coupled scalar field, the gravitational couplings come from the Lagrangian
\beqa
\sqrt{-g} {\cal L}&=& \frac{\sqrt{-g}}{2}\left[g^{\mu\nu}\partial_\mu \phi \partial_\nu \phi - m^2\phi^2\right] \nonumber \\
&=& \frac12 \left[\left(1+\frac{h}{2 } + \frac{h^2}{8}-\frac12h^{\alpha\beta}h_{\alpha\beta}\right)\left(\partial_\mu \phi \partial^\mu \phi -m^2\phi^2\right)-h^{\mu\nu}\partial_\mu \phi \partial_\nu \phi \right. \nonumber \\
&~&\left. ~~~+\left(h^{\mu\beta}h^{~\nu}_\beta - \frac12  h h^{\mu\nu}\right)\partial_\mu \phi \partial_\nu \phi\right] 
\eeqa
Here $h= h^\lambda_\lambda$.

\begin{figure}[htb]
\begin{center}
\includegraphics[height=30mm,width=80mm]{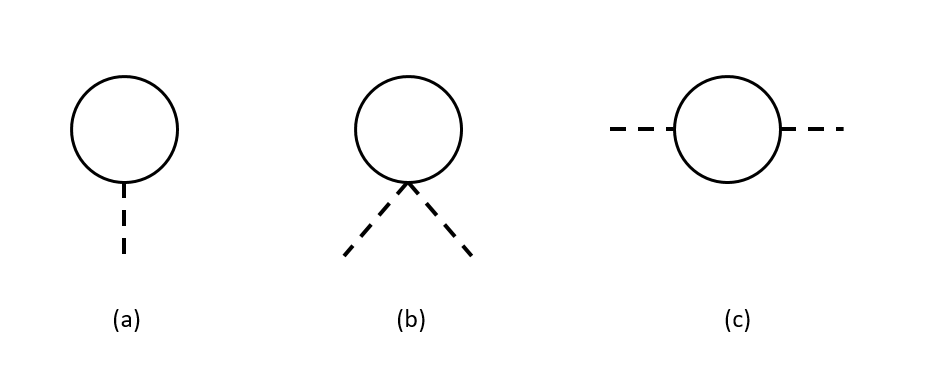}
\caption{The tadpole diagrams (a) and (b) and the bubble diagram (c). The solid line is the scalar field and the dashed line is the metric field. }
\label{diagrams}
\end{center}
\end{figure}

The one loop correction linear in the gravitational field comes from the tadpole diagram of Fig \ref{diagrams}a, and has the form 
\beqa
-i{\cal M}_a&=& -i h^{\mu\nu} \int \frac{d^4k}{(2\pi)^4} ~k_\mu k_\nu \frac{i}{k^2-m^2+i\epsilon} \nonumber \\
&=& -i h^{\mu\nu} \mu^{4-d}\int \frac{d^dk}{(2\pi)^d} ~\frac{\eta_{\mu\nu}}{d} k^2\frac{i}{k^2-m^2+i\epsilon} 
\eeqa
where I have dropped a dimensionless integral because I am using dimensionless regularization\footnote{The effect of this dimensionless integral also vanishes when using a cutoff because there is an extra interaction from the measure required when using cutoff regularization \cite{Donoghue:2020hoh} .}. This integral is readily evaluated and yields
\beq
-i {\cal M}_a = i h \frac{m^4}{64\pi^2} \left[ \frac{1}{\bar{\epsilon}} +\log \frac{\mu^2}{m^2} +\frac32 \right] 
\eeq
with the shorthand notation
\beq
\frac{1}{\bar{\epsilon}} = \frac{1}{\epsilon} -\gamma+\log(4\pi) \  \ .
\eeq

The second order terms in the gravitational field are similarly straightforward. That of Figure \ref{diagrams}b is given by
\beqa
-i{\cal M}_b&=& i \left(h^{\mu\beta}h_\beta^{~\nu} -\frac12 h h^{\mu\nu}\right) \int \frac{d^4k}{(2\pi)^4} ~k_\mu k_\nu \frac{i}{k^2-m^2+i\epsilon}\nonumber   \\
&=& -i\frac{m^4}{ 64\pi^2}\left(h^{\mu\beta}h_\beta^{~\nu} -\frac12 h h^{\mu\nu}\right)\eta_{\mu\nu} \left[ \frac{1}{\bar{\epsilon}} +\log \frac{\mu^2}{m^2} +\frac32 \right] \ \ .
\eeqa
For the bubble diagram of Figure \ref{diagrams}c, for our purposes in this section we can evaluate it at zero momentum because the cosmological constant does not involve momentum dependence. This yields
\beqa\label{bubble}
-i{\cal M}_c&=&  \frac12 \int \frac{d^4k}{(2\pi)^4} \frac{ \left[ h^{\mu\nu} k_\mu k_\nu -\frac12 h(k^2-m^2)\right]\left[ h^{\alpha\beta} k_\alpha k_\beta -\frac12 h(k^2-m^2)\right]}{(k^2-m^2+i\epsilon)^2} \nonumber \\
&=&\frac12 \int \frac{d^4k}{(2\pi)^4}\left[ h^{\mu\nu} h^{\alpha\beta}  \frac{  k_\mu k_\nu k_\alpha k_\beta}{(k^2-m^2+i\epsilon)^2} - h h^{\mu\nu} \frac{k_\mu k_\nu}{k^2-m^2+i\epsilon}\right] \nonumber \\
&=&   \frac12 \int \frac{d^4k}{(2\pi)^4}\left[\frac{ h^{\mu\nu} h^{\alpha\beta} }{d(d+2)}(\eta_{\mu\nu} \eta_{\alpha\beta} +\eta_{\mu\alpha}\eta_{\nu\beta}+ \eta_{\mu\beta}\eta_{\nu\alpha})\left( \frac{   m^4}{(k^2-m^2+i\epsilon)^2} + \frac{2m^2}{k^2-m^2+i\epsilon}\right) \right.  \nonumber \\
&~& \left. ~~~~~~~~~~~~~~~~~~~~~~~~~~~~ -\frac{h h^{\mu\nu} }{d}\frac{m^2}{k^2-m^2+i\epsilon}\right] 
\nonumber \\
&=&i\frac{m^4}{ 128\pi^2}\left(h^{\mu\nu}h_{\mu\nu} -\frac12 h^2 \right) \left[ \frac{1}{\bar{\epsilon}} +\log \frac{\mu^2}{m^2} +\frac32 \right] \ \ .
\eeqa
In this case, it is non-trivial that the finite part (the $3/2$ within the final square bracket) comes out identical to the finite part in the previous two diagrams. However, this is required for general covariance. Summing the three diagrams one gets the appropriate combination for the expansion of $\sqrt{-g}$, i.e.
\beq
-i{\cal M}= i\frac{m^4}{ 32\pi^2}\left(\frac12 h +\frac 18 h^2 -\frac14 h^{\mu\nu}{\mu\nu} \right) \left[ \frac{1}{\bar{\epsilon}} +\log \frac{\mu^2}{m^2} +\frac32 \right] \ \ . 
\eeq
This lets us read off the one-loop contribution to the cosmological constant
\beq\label{deltaLambda}
\delta \Lambda = -\frac{m^4}{32\pi^2} \left[  \frac1{\epsilon} -\gamma +\log(4\pi) +\log \frac{\mu^2}{m^2} +\frac32 \right] \ \ . 
\eeq

This quantum effect is one contribution to the physical renormalized value of the cosmological constant. All physical predictions are expressed in terms of the measured value after renormalization. As is well known, it is striking that the measured value is so much smaller than all the known mass scales. The present calculation reinforces that mystery, but does not help explaining it. 

One can also readily calculate the renormalization of the Einstein term, proportional to $R$, by this method. In this case one must include the external momentum dependence. The result is proportional to $m^2$, and does not add anything particular to our discussion, so I do not describe it further. 

\section{The non-local partner}

While the ingredients calculated in the previous section disappear into the renormalized value of $\Lambda$, non-local effects are finite and physical. Having worked through the renormalization calculation, we can readily see that such effects must occur. The fact that the scalar bubble diagram contributes to the renormalization, as seen in Eq. \ref{bubble}, and that this diagram has logarithmic momentum dependence, as seen in Eq. \ref{scalarbubble}, mean that there will be logarithmic non-localities when one includes non-zero momentum/spatial dependence. 

For this calculation, one needs to include the external momentum dependence in the bubble diagram, which will be denoted by the momentum $q^\mu$. The reduction to scalar integrals is more extensive in this case. It is however useful to show one such integral which displays an important facet of the final result. Consider the integral
\beqa\label{fourmomentaintegral}
I_{\mu\nu\alpha\beta} &=&  \int \frac{d^4k}{(2\pi)^4} \frac{k_\mu k_\nu k_\alpha k_\beta}{[k^2 -m^2+i\epsilon]  [(k+q)^2 -m^2 +i\epsilon]}  \nonumber \\
&=& F\left( \eta_{\mu\nu}\eta_{\alpha\beta} +   \eta_{\mu\alpha}\eta_{\nu\beta} + \eta_{\mu\beta}\eta_{\alpha\nu} \right). \nonumber \\
&+&G\left(  \eta_{\mu\nu} q_\alpha q_\beta + \rm{ 5~ perms.}  \right) \nonumber \\
&+& H q_\mu q_\nu q_\alpha  q_\beta \ \ .
\eeqa
The $m^4$ portions of the coefficient functions $F,~G,~H$ are given by
\beqa
F&=& \frac{i}{16\pi^2} \frac{m^4}{8}\left[ \frac{1}{\bar{\epsilon}} +\log \frac{\mu^2}{m^2} +\frac32 -\frac{8}{15}J(q^2)\right] + {\cal O}(m^2)\nonumber \\
G&=& \frac{i}{16\pi^2} \frac{m^4}{15q^2}  J(q^2) + {\cal O}(m^2)\nonumber \\
H &=& \frac{i}{16\pi^2} \frac{-m^4}{5q^4} J(q^2) + {\cal O}(m^2)\eeqa
where $J(q^2)$ is the logarithm of Eq. \ref{Jintegral}. The feature worth noticing is the inverse powers of $q^2$ in the $G$ and $H$ formfactors. Combined with the numerator factors in the integral definition of Eq. \ref{fourmomentaintegral}, these mean that all of these terms are of zeroth order when counting powers of the momenta. 
 
 Consider a matrix element with two external gravitons having polarization vectors $\epsilon_{\mu\nu}$ and $\epsilon^*_{\alpha\beta}$ and define
\beq
 Q_{\mu\nu}= q_\mu q_\nu -\eta_{\mu\nu} q^2
 \eeq
 with $q$ being the momentum carried by the gravitons. With this notation, the residual matrix element at order $m^4$ and $m^2$ after renormalization has been performed is
\beq\label{matrixanswer}
{\cal M}_{\mu\nu\alpha\beta} =\frac{1}{160\pi^2 q^4}\left(Q_{\mu\nu}Q_{\alpha\beta} + Q_{\mu\alpha}Q_{\nu\beta} + Q_{\mu\beta}Q_{\nu\alpha}\right)\left[ m^4 J(q^2) +\frac16 m^2 q^2  - 3m^2q^2 J(q^2)  \right]
\eeq
There are also terms of order $m^0$ , which contribute to the Barvinsky Vilkovisky action in ways which are described above. What is most important to note is the overall factor of $1/q^4$. Combined with the factors of $q$ in the $Q_{\mu\nu}$, this makes the overall pre-factor zeroth order in the momentum expansion.

In position space this interaction can only be represented  by a non-local effective action. This can be first displayed using the gravitational field $h_{\mu\nu}$ as a preliminary step. The gravitational field is not gauge invariant by itself, and in this case the choice of harmonic gauge, $\partial_\mu h^{\mu\nu} =\frac12 \partial^\nu h$ will get rid of the pesky inverse powers of $q^2$. In this gauge, we have
\beqa
 \epsilon^{\mu\nu} Q_{\mu\nu} &=& -\frac12 q^2 \epsilon^\lambda_{~\lambda} \nonumber \\
 \epsilon^{\mu\nu} Q_{\mu\alpha} &=& \frac12 q^\nu q_\alpha \epsilon^\lambda_{~\lambda} -q^2 \epsilon^\nu_\alpha
 \eeqa
 In terms of the field, and again using the quasi-local notation, we find
 \beq
 {\cal L} =\frac{m^4}{160\pi^2 } \left[h_{\mu\nu} \log[(\Box +m^2)/m^2 ]h^{\mu\nu}. -\frac18 h  \log[(\Box +m^2)/m^2 ]h \right]
  \eeq
where we have defined the notation $ \log [(\Box+m^2/m^2)]$  by
\beq
\langle x|  \log [(\Box+m^2/m^2) | y\rangle = \int \frac{d^4q}{(2\pi)^4} e^{iq\cdot(x-y)} J(q^2) \  \ .
\eeq
More discussion of this function is found in the appendix.

In turning this into a covariant expression, we need to provide a correspondence between the possible curvatures and their matrix elements. At second order in the fields the different products of curvatures have matrix elements.
\beqa
\langle RR\rangle &=& 2Q_{\mu\nu} Q_{\alpha\beta}  \nonumber \\
\langle R_{\lambda\sigma}R^{\lambda\sigma} \rangle &=&  \frac14 \left[2Q_{\mu\nu} Q_{\alpha\beta} + Q_{\mu\alpha} Q_{\nu\beta} + Q_{\mu\beta} Q_{\nu\alpha}\right]
\eeqa
When using the curvatures, we still have the inverse powers of $q^2$ in the matrix element so we use
\beq
\langle (\frac1{\Box} R)  (\frac1{\Box} R) \rangle = \frac{2}{q^4} Q_{\mu\nu} Q_{\alpha\beta} 
\eeq
This leads to the non-local action
\beqa
{\cal L} &=&  \frac{m^4}{40\pi^2}\left[ \left(\frac1{\Box}R_{\lambda\sigma}\right)\log((\Box+m^2)/m^2)\left(\frac1{\Box}R^{\lambda\sigma} \right)-\frac18 \left(\frac1{\Box}R\right) \log((\Box+m^2)/m^2)\left(\frac1{\Box}R\right) \right] \nonumber \\
&+&  \frac{m^2}{240\pi^2} \left[R_{\lambda\sigma}\frac{1}{\Box}R^{\lambda\sigma} -\frac18R\frac{1}{\Box}R \right]  
\eeqa
Here I have included the effect of the second term in the square brackets of Eq. \ref{matrixanswer} even though it is second order in the derivative expansion, as it will be important for the discussion of decoupling and locality which will be discussed soon. The first line is the non-local partner of the cosmological constant.

\section{Decoupling}

At first sight, this operator would seem to violate the principle of decoupling \cite{Appelquist:1974tg}. The effects of very heavy masses should be local, when applied at energies much lower that the mass. The factors of $1/\Box $ seem to violate this as they appear non-local down  to low energies. However, this is not the case. The power counting is different at low and high energy because
\beqa
J(q^2) &\sim&\log \frac{-q^2}{m^2}-2~~~~~~~~   q^2\gg m^2.  \nonumber \ \\
&\sim& - \frac{q^2}{6m^2}~~~~~~~~~~~~~q^2\ll m^2
\eeqa
This implies that at low energy there is a cancelation between the terms in Eq. \ref{matrixanswer}. The overall factor in the square brackets of Eq. \ref{matrixanswer} is of order $q^4$ for $q^2\ll m^2$ and results in a local operator. 

This implies that the non-local operator is only operative at scales above the particle mass. What is meant by the phrase ``scales'' depends on the context. In the original computation, this clearly referred to the momentum. However in general relativistic applications, it would refer rather to spatial or temporal derivatives. In cosmological settings, the Hubble parameter $H= \dot{a}/a$ would likely play the role of the energy scale in applications. 

The presence of the integral $J(q^2)$ has been previously noted in various settings in gravitational physics\cite{Dalvit:1994gf, Gorbar:2002pw, Burns:2014bva} including appropriate comments about non-localities and decoupling. These works provide possible avenues for application of the effective action described above. 

\section{Discussion}

I have calculated the non-local effects due to quantum loops of a massive particle, and found an effect which is zeroth order in the derivative expansion, much like the cosmological constant. Non-local effects are ubiquitous when loops involve particles whose mass is smaller than the relevant energy scales. In this case the result is special because it is very many orders of magnitude larger than the local effect (the cosmological constant), although the non-locality is active only at high energy scales\footnote{A related special case is the inflaton potential, where the local component can be fine-tuned to be almost flat. However, Miao and Woodard\cite{Miao:2015oba} have pointed out that couplings of the inflaton to fields needed for reheating generate a non-local component which cannot be fine-tuned.}. I have used the Barvinsky-Vilkovisky technique to express this in covariant form involving the curvatures. 

The cosmological constant is not a running coupling constant in the usual sense. Its value in applications does not depend on the energy scales involved. One might be tempted to identify the unphysical dimensional regularization parameter $\mu$, which appear in the renormalization procedure of  Eq. \ref{deltaLambda}, as a running parameter. In mass-independent renormalization schemes, or when the mass is negliable, this $\mu$ dependence is sometimes used as a surrogate for the energy dependence, because in those theories the scale dependence tracks the energy dependence. However, here it just totally disappears into the renormalized parameter and does not track any energy dependence, as can be seen most obviously in the tadpole loop. Energy dependence in the coupling could be described in position space by a non-local effective action\footnote{For an example with QED, see \cite{Donoghue:2015nba}}. In the case of the cosmological constant, the non-local effect has a different structure than the local effect. The non-local partner is the closest thing that we have to a running effect at zeroth-order in the derivative expansion. 

The triangle diagram can bring in further non-local effects. Some of these will be part of the interaction calculated in this paper, when the curvature products are expanded to order $h^3$. There likely will also be new operators which can be best expressed as the product of three curvatures. Some of these can also be of zeroth-order in the derivative expansion. The result of this paper is an approximation to a much more complicated general result. This expansion in the curvatures is a weak field expansion. It is also possible that repeating this procedure around other fixed background spacetimes may yield forms which are more useful for those backgrounds. 

There is a subset of the literature dealing with non-local actions in gravity, which can be traced through the references in \cite{Barvinsky:2003kg, Deser:2007jk, Deser:2019lmm, Maggiore:2014sia,  Maggiore:2016gpx}. Many of these are speculative suggestions for non-local actions. In contrast, the Barvinsky Vilkovisky program and the present work are derived from the quantum effects of standard local theories and are not speculative. 

Techniques in applying non-local actions are less well-developed than those of local Lagrangians, and further work here would be useful for applications.

\section*{Acknowledgements}  This work has been partially supported by the US National Science Foundation under grant NSF-PHY-21-12800.
I thank Basem El-Menoufi, Roberto Percacci and Richard Woodard for discussions.  

\section{Appendix - Covariant non-local functions}

In flat space the meaning of $\log \Box$ is not controversial - its matrix element is the Fourier transform of $\log (-q^2-i\epsilon)$ (Eq. \ref{logbox}). Notationally it appears as an operator, but it is applied as just a function. When we generalize to curved spacetime, we seek a covariant definition which reduces to the flat space version in the appropriate limit. Such a generalization will involve the metric as well as the endpoints. Unfortunately there is not a unique choice - I display three possibilities below and there are undoubtably more. The non-local operators $\log \Box$ and $\log (\Box+\frac16 R)$ also are discussed in connection with the conformal anomaly \cite{Deser:1976yx, Deser:1993yx, Deser:1999zv} as these provide a representation for the anomaly.

Given this non-uniqueness, how can one proceed? A first point is that within the framework of an expansion in the curvatures differences in the definitions can be corrected for by adjusting higher order terms in the curvature expansion. If you are working at second order in the curvature, the differences between the various covariant operators appears at third order in the curvature. At leading order, one may in principle use any covariant definition. 

However, it is still possible that some forms are better than others for particular applications. For example, a given form may introduce spurious infrared divergences which should not be present. This has been seen in at least one application \cite{Donoghue:2015nba}. Beyond this caveat, not much is known.

Here are three possible definitions of covariant $\log \Box$:

\noindent 1) Barvinsky Vilkovisky and collaborators use a definition with a single propagator,
\beq
\langle x|\log \Box/\mu^2 |y> = \int_0^\infty dm^2 \langle x|\left[\frac1{\Box - m^2}-\frac1{\mu^2-m^2}\right] |y\rangle  \ \ .
\eeq
Here 
\beq
\langle x|\frac1{\Box - m^2}|y\rangle
\eeq
is the covariant propagator for a massive particle \cite{DeWitt:1975ys}. 

\noindent 2) Osborn and Erdmenger have proposed using two massless propagators, and the subtraction of the divergence which occurs in this combination
\beq
\langle x|\log \Box/\mu^2 |y> = \left[ \langle x|\frac1{\Box - m^2}|y\rangle\right]^2 - \frac1{\sqrt{-g}} \delta^4(x-y) \left[\frac1{\bar{\epsilon}} +\log \mu^2/m^2 +1\right] \ \ .
\eeq
The flat space version of this is the scalar bubble diagram, so this definition has the good feature that it close to what one does when calculating Feynman diagrams. One proves this relation by Fourier transforming both sides.

\noindent 3) Perhaps the simplest generalization uses the proper time parameterization. In this case one uses the flat space representation
\beqa
L_0(x-y) &=& \langle x|\log \Box/\mu^2 |y\rangle \nonumber \\
&=&\int \frac{d^4q}{(2\pi)^4}e^{-iq\cdot (x-y)}\log ( -q^2-i\epsilon)/\mu^2  \nonumber \\
&=& \int \frac{d^4q}{(2\pi)^4} ~\int_{0}^{\infty}  \frac{ds}{s}   \left[e^{i(q^2+i\epsilon)s}- e^{i\mu^2 s}\right]  e^{-iq\cdot (x-y)}  \nonumber \\
&=& \frac1{16\pi^2} \int_{0}^{\infty} \frac{ds}{s^3} e^{-i \frac{(x-y)^2 -i\epsilon}{4s}}` +~ {\rm local}
\eeqa
This can be made covariant most simply by replacing 
the space time separation by the geodetic distance, i.e. the proper distance along the geodesic connecting the points $\frac12 (x-y)^2 \to \sigma$ so that one can use
\beq
L(\sigma) = \frac1{16\pi^2} \int_{0}^{\infty} \frac{ds}{s^3} e^{-i \frac{\sigma -i\epsilon}{2s}}
\eeq

That these definitions are different can be verified by expanding them about flat spacetime in an expansion in the gravitational field.

There is another aspect to the use of non-local functions, which is the fact that tensor quantities transform differently at different spacetime locations because general coordinate invariance is a local symmetry. In $R_{\mu\nu}(x) \langle x|\log \Box |y\rangle^{\mu\nu\alpha\beta} R_{\alpha\beta}(y)$ the curvature tensors transform the differently at $x$ and $y$. The treatment of this using Wilson lines is reserved for a future publication \cite{nonlocalBasem}. 

The function
\beq
J(x,y) = \langle x| \log (\Box +m^2)/m^2 |y\rangle
\eeq
is a covariant generalization of 
\beq
J_0 (x-y) = \int \frac{d^4q}{(2\pi)^4} e^{iq\cdot(x-y)} J(q^2) = \int \frac{d^4q}{(2\pi)^4} e^{iq\cdot(x-y)} \int_0^1 dx ~\log\left[\frac{m^2 -x(1-x)q^2}{m^2} \right]  \  \ .  \ .
\eeq
Again, this is not uniquely defined. Because it is to be applied only above the energy scale $m$, it is probably most useful to use an approximation valid for $q^2 \gg m^2$, namely $J(q^2) \sim \log( -q^2/m^2) -2$, in which case the non-local part of $J(x,y)$ is approximated by $\log \Box$.

 \end{document}